\documentclass[a4paper,11pt]{article}
\usepackage[utf8]{inputenc}
\pdfoutput=1
\usepackage{multirow}
\usepackage{jheppub}
\usepackage{subfig}
\usepackage{tikz}
\usepackage{setspace}
\allowdisplaybreaks

\title{Boosting Assisted Annihilation for a Cosmologically Safe MeV Scale Dark Matter}

\author[a]{Ujjal Kumar Dey,}
\author[b]{Tarak Nath Maity,}
\author[b, c]{Tirtha Sankar Ray}
\affiliation[a]{Asia Pacific Center for Theoretical Physics,\\
				Pohang 37673, Korea}
\affiliation[b]{Department of Physics, Indian Institute of Technology Kharagpur,\\
Kharagpur 721302, India}
\affiliation[c]{Centre for Theoretical Studies, Indian Institute of Technology Kharagpur,\\
Kharagpur 721302, India}

\emailAdd{ujjal@apctp.org}
\emailAdd{tarak.maity.physics@gmail.com}
\emailAdd{tirthasankar.ray@gmail.com}

\abstract{
Assisted annihilation generates thermal sub-GeV dark matter through a novel annihilation between a pair of dark matter and standard-model-like states, called the “assister”. We show that, depending on the mass hierarchy between the assister and dark matter, there can be either a suppression or a boost of the effective cross section. This augmentation enables the possibility of $\mathcal{O}(100)$ MeV scale dark matter with perturbative coupling that saturates the relic density estimates while being relatively insulated from cosmological constraints like big bang nucleosynthesis and cosmic microwave background.
}

\preprint{\begin{flushright}
APCTP Pre2018 - 021
\end{flushright}}

\begin{document}

\maketitle
\flushbottom

\section{Introduction}
\label{sec:intro}
The landscape of particulate interpretation of dark matter (DM) is dominated by the weakly interacting massive particles (WIMPs) that can explain the astrophysically measured dark matter abundance. Typically they have   masses in the GeV scale with weak scale two-body annihilation cross sections driving their freeze-out in the early Universe. The WIMPs, which fit well within the $\Lambda$CDM framework, have been increasingly constrained from a synergy of non-observation in direct and indirect detection experiments.  A growing  conundrum in the mismatch  between observations and simulations at the galactic scale structure of the Universe \cite{Nakama:2017ohe, Bullock:2017xww} provides a motivation for   sub-GeV light dark matter (LDM) with self-interaction~\cite{Spergel:1999mh}. An $N (\geq 3)\rightarrow 2$ process driving thermal freeze-out, can naturally lead to LDM~\cite{Dolgov:1980uu, Dolgov:2017ujf, Hochberg:2014dra, Hochberg:2014kqa, Dey:2016qgf, Bernal:2015xba, Lee:2015uva, Choi:2015bya, Bernal:2015bla, Hochberg:2015vrg, Kuflik:2015isi, Choi:2016tkj, Choi:2016hid, Bernal:2017mqb, Ho:2017fte, Cline:2017tka, Choi:2017zww, Kuflik:2017iqs, Hochberg:2018rjs, Hochberg:2018vdo}. This type of annihilation process can be relevant if (i) the 2-body cross section is forbidden or strongly  suppressed, (ii) densities of the annihilating species are appreciable, and (iii) the velocity of the interacting particles are low~\cite{Namjoo:2018oyn}. One novel example of this class of models is the \textit{assisted annihilation}, where, along with DM particles, there are standard model(SM)-like \textit{assisters} in the initial state facilitating the annihilation of LDM~\cite{Dey:2016qgf}. 

In this paper, we  make the first systematic study of the cosmological consequences  of the assisted annihilation mechanism in light of stringent constraints on LDM from relic density measurements,  big bang nucleosynthesis (BBN) and cosmic microwave background (CMB). The $4\to 2$  or higher-order annihilation scenario with perturbative couplings implies a keV or lighter scale DM,  which  are strongly constrained from CMB observations. Thus, we focus on  $3\to 2$ assisted annihilation topology that can lead to MeV scale DM. Crucially, the flexibility accorded by the set-up allows the mass hierarchy between the DM and the assister to regulate the Boltzmann factor, resulting in  a suppression  or  a novel boost of the effective thermally averaged annihilation cross section. We show that the allowed mass range for DM can easily extend to hundreds of MeV while maintaining constraints from relic density and perturbativity. This may be contrasted with the usual $3\to 2$ annihilation paradigm  for DM \cite{Dolgov:1980uu, Dolgov:2017ujf, Hochberg:2014dra, Hochberg:2014kqa, Bernal:2015xba, Cline:2017tka} where in the absence of a  boost the   perturbativity considerations put severe upper limits on the DM mass, other than tuned resonance effects \cite{Choi:2016hid}.

The presence of  a relatively light DM and  assister can alter the primordial light elements' abundances  causing them to deviate from BBN observations. The consonance in the baryon to photon ratio ($\eta$) measurement from BBN and CMB and the $N_{\rm eff}$ measurements from CMB add additional constraints. We find that the  relative  mass hierarchy of the assisters and the DM states  together with the decay width of the assisters determine the viable windows for this scenario.  
%

\section{Effective Parametrization}
\label{sec:setup}
\begin{figure}[t]
\begin{center}
\subfloat[\label{sf:asstann}]{
\includegraphics[scale=0.6]{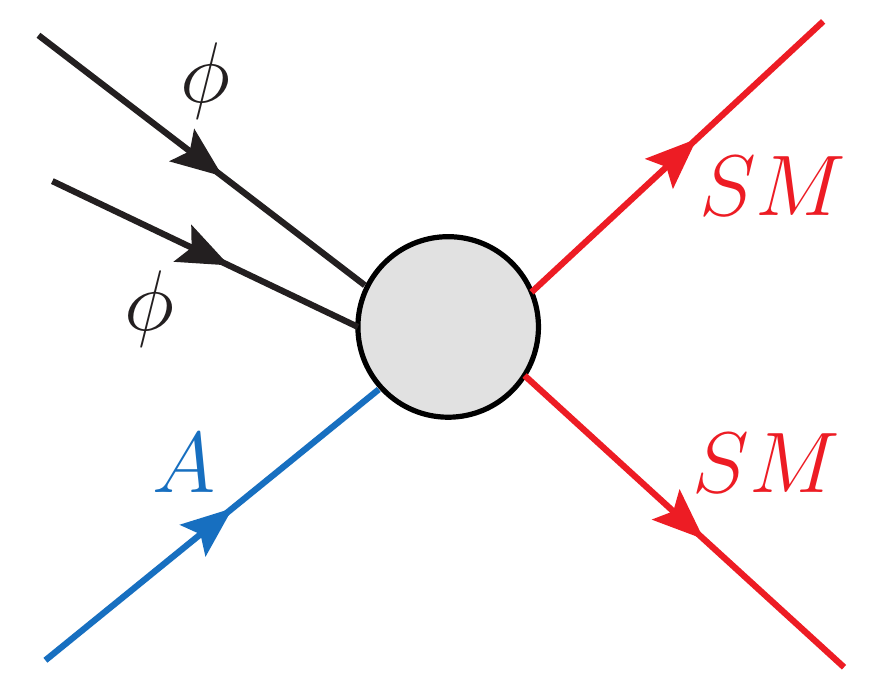}}~~
\subfloat[\label{sf:nbolt}]{
\includegraphics[scale=0.6]{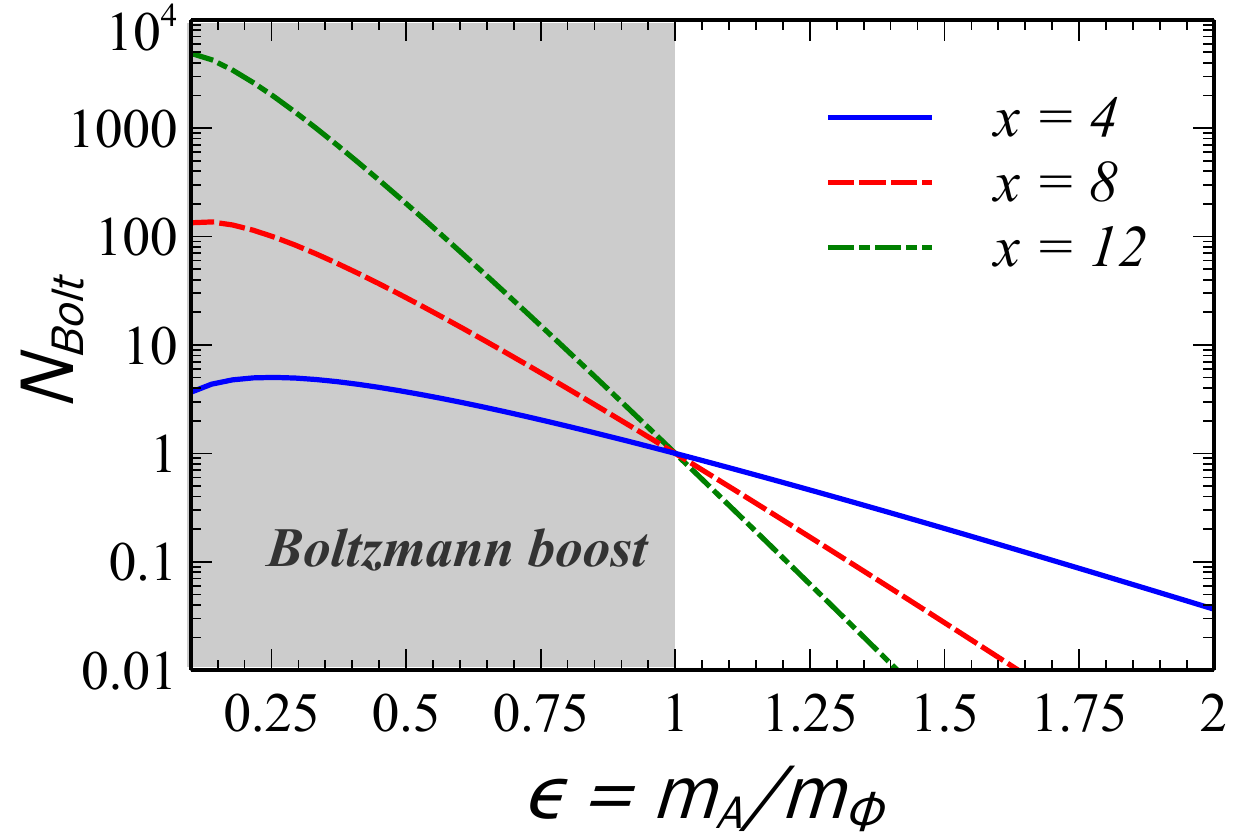}}
\caption{(a) Block diagram of a typical $3\to 2$ assisted annihilation channel. (b) Boltzmann factor $N_{\rm Bolt}$ as a function of mass ratio between assister and DM for various $x$ with $3 \to 2$ assisted annihilation.}
\label{fig:nboltfreez}
\end{center}
\end{figure}

Once it is pared down, this framework includes a stable particle $\phi$ which is the DM candidate and an assister $A$ that can decay to SM states. After their respective chemical decoupling the number density of assisters starts depleting  while the DM states freeze out. Within the framework of real scalar fields the dominant assisted annihilation topology is for a pair of DM and a single assister to annihilate to  SM or SM-like states as depicted in figure~\ref{sf:asstann}.  This can arise in scenarios where  lower-order  $2 \to 2$ annihilations of the DM are  suppressed. 
Remaining agnostic with the details of the model, we can parametrize the relevant thermally averaged cross sections as,
\begin{subequations}
\begin{gather}
\langle \sigma v^{2} \rangle_{\phi \phi A \to \text{SM}~\text{SM}} = \frac{\alpha^2_1}{m_{\phi}^5}, \label{se:aaXsection}\\
\langle \sigma v \rangle_{A A \leftrightarrow \phi \phi } = \frac{\alpha^2_2}{m_{\phi}^2},~~
\langle \sigma v \rangle_{A A \to \text{SM}~\text{SM}} = \frac{\alpha^2_3}{m_{A}^2}~,
\end{gather}
\label{eq:effect-para}
\end{subequations}
where $\alpha_{1,2,3}$ are the corresponding effective couplings while $m_{\phi}$ and $m_{A}$ are the masses of DM and assister respectively. Assuming  $\alpha_1 \gg \alpha_{2}$ the assisted annihilation process will dominate the freeze-out of the DM. In equation \eqref{se:aaXsection} the final states are assumed to be heavy SM or SM-like states so as to prevent the possibility of the  $2 \to 3$ ($\phi \phi \to A\, \text{SM}~\text{SM}$) channel to open up and overwhelm the assisted annihilation process. However these states can subsequently decay to light SM states and a specific realization based on the same will be discussed later in this article. Note that the DM relic density is relatively independent of $\alpha_{3}$ and the assister decay rate $\Gamma_{A\to \text{SM}~\text{SM}}$. 

\section{Relic Density}
\label{sec:relic}
%
The rate of the assisted annihilation at any instant depends on the number density of DM particles as well as assisters. The present day relic abundance of the DM is obtained by tracking the evolution of number densities of both the DM particles and assisters and can be compared with experimental observations~\cite{Aghanim:2018eyx}. This mandates a numerical solution of the coupled Boltzmann equations involving number densities of both the species. The coupled Boltzmann equations in terms of co-moving number densities of DM ($Y_{\phi}=n_{\phi}/s$) and assister ($Y_{A}=n_{A}/s$) in terms of the parametrization given in equation \eqref{eq:effect-para} can be written as
%
%
\begin{subequations}
\begin{align}
\label{seq:boltzYY}
\frac{dY_{\phi}}{dx} = & -\frac{x s^2 g^{1/2}_{*}}{H}
                        N_{\rm Bolt}
                        \langle \sigma v^{2} \rangle_{\phi \phi A
                         \to \text{SM}~\text{SM}}
                        \left(Y_{\phi}^{2} Y_{A} \frac{
                        Y_{\phi}^{\rm eq}}{Y_{A}^{\rm eq}}
                         - \left(Y_{\phi}^{\rm eq}\right)^{3}
                        \right) + 
                        \frac{x s g^{1/2}_{*}}{H}
                         \langle \sigma v \rangle_{A A 
                         \leftrightarrow \phi
                         \phi} \nonumber \\
                       & \times  \left\lbrace
                         \Theta(m_A - m_\phi)
                         \left(Y^2_{A}-\left(
                         \frac{Y_{\phi}Y^{\rm eq}_A}
                         	  {Y^{\rm eq}_{\phi}} \right)^2 \right)-
                         \Theta(m_\phi - m_A) 
                         \left(Y^2_{\phi}-\left(
                          \frac{Y_{A}Y^{\rm eq}_{\phi}}
                          	   {Y^{\rm eq}_{A}}
                          	   \right)^2\right) \right\rbrace, \\
\frac{dY_{A}}{dx} = & -\frac{x s g^{1/2}_{*}}{H}
					   \langle \sigma v 
					   \rangle_{A A \leftrightarrow \phi \phi}
					   \nonumber \\
					& \times \left\lbrace
					    \Theta(m_A - m_\phi)
                        \left(Y^2_{A} - \left(
                          \frac{Y_{\phi} Y^{\rm eq}_A}
                               {Y^{\rm eq}_{\phi}}
                               \right)^2\right)-
                        \Theta(m_\phi - m_A)
                        \left(Y^2_{\phi} -  \left(
                          \frac{Y_{A}Y^{\rm eq}_{\phi}}
                               {Y^{\rm eq}_{A}}
                               \right)^2\right) \right\rbrace  
                       \nonumber \\
                    & -\frac{x s g^{1/2}_{*}}{H}
                       \langle \sigma v 
                       \rangle_{A A \to \text{SM}~\text{SM}} 
                       \left \lbrace 
                         Y^2_A - \left(Y^{\rm eq}_A\right)^2
                       \right \rbrace - 
                       \frac{g^{1/2}_{*}\Gamma_{A\to \text{SM}~\text{SM}}} 
                            {x H} 
                       \left(Y_A-Y^{\rm eq}_A\right), \\
\mbox{with} ~~ N_{\rm Bolt} & = e^{x(1-\epsilon)}\epsilon^{3/2},
              ~ g^{1/2}_{*} = 1 + \frac{1}{3}
              					  \frac{d(\text{ln}~g_s)}
              					       {d(\text{ln}~T)}~.
\end{align}
\end{subequations}
%
%
In these expressions $x=m_{\phi}/T$, $\epsilon=m_{A}/m_{\phi}$, entropy density $s=2 \pi^2 g_s T^3/45$, Hubble constant $H= \sqrt{\pi^2 g_{\rho}/90}\left(T^2/M_{\rm Pl}\right)$, and $g_s$ and $g_{\rho}$ are the effective number of relativistic degrees of freedom corresponding to entropy and energy density, respectively. In these equations, we have taken into account the temperature dependence of $g_s$ and $g_{\rho}$, and consequently of $g^{1/2}_{*}$~\cite{Drees:2015exa}.
The parameter $N_{\rm Bolt}$ modulates the annihilation cross section of the process and physically it reflects the fact that  assisted annihilation can continue as long as the number densities of both the DM and the assister are appreciable in the early Universe. The variation of $N_{\rm Bolt}$ against the mass ratio of assister and DM, $\epsilon$, has been depicted in figure~\ref{sf:nbolt}. From this figure, it is evident that either there will be a boost or a suppression depending on mass hierarchy between DM and assister.  A systematic discussion of both  scenarios is now in order.
%

\subsection{Case I ($m_{\phi} \leq m_{A}$)}
\label{sbsc:case1}
This is the typical case where the DM is the lightest particle in the initial state and can be viewed as a generalization of the co-annihilation topology. With the evolution of the Universe, the equilibrium co-moving number density of a non-relativistic species falls as $\sim \text{exp}(-m/T),$ where $m$ is the mass of the species. This implies that the lower the value of $m$, the larger the number density of the species for a given temperature. Therefore, for $m_{\phi} < m_A$, a smaller number of assisters are available in the thermal soup to interact with the DM during freeze-out. This leads to a suppression in the effective cross section through $N_{\rm Bolt}$ as depicted in figure~\ref{sf:nbolt}.
With this mass hierarchy, a  successful freeze-out driven by assisted annihilation within the  perturbative limit,  would  require DM and assister  masses to be relatively degenerate. In figure~\ref{fig:relic} we show the relic-density-allowed contour for $m_{A}/m_{\phi}=1.5$ in the  $m_{\phi}$ vs. $\alpha_1$ plane by the green dashed line. A larger value of  $\epsilon$ will gradually  get relegated to the strong coupling region. The maximum allowed DM mass of $\sim 20$ MeV  for the conservative limit of $\alpha_1 \leq 1$ is obtained for $\epsilon =1.$
%

\subsection{Case II ($m_{\phi} > m_{A}$)}
\label{sbsc:case2}
\begin{figure}[t]
\begin{center}
\includegraphics[scale=0.35]{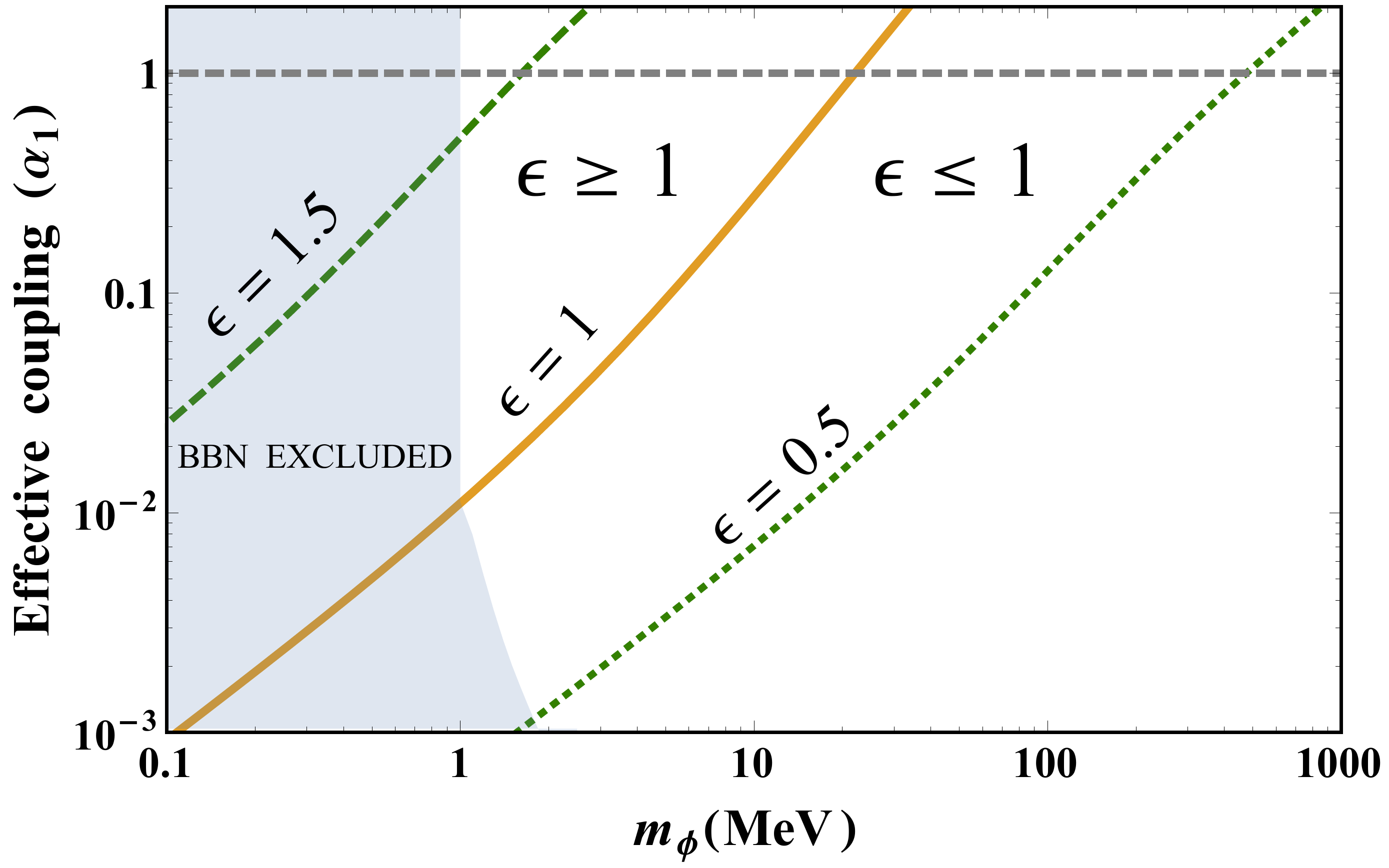}
\caption{Relic-density-allowed parameter space in $m_{\phi}-\alpha_1$ plane. The orange line represents relic-density-allowed contour for $\epsilon = 1$, where $\epsilon \equiv m_A/m_{\phi}$. The region above and below the $\epsilon = 1$ line corresponds to a  Boltzmann suppression ($\epsilon > 1$), or boost ($\epsilon < 1$) respectively. The light blue shaded region shows the BBN excluded parameter space. The gray dashed line shows the $\alpha_1 = 1$ line.}
\label{fig:relic}
\end{center}
\end{figure}

One of the distinguishing features of the assisted annihilation set-up is that the DM and the assister are not charged under the same stabilizing symmetry. Thus, in contrast to the usual co-annihilation scenario, it is not mandatory for DM to be lighter than the assister.  With $m_{\phi} > m_{A}$ the lighter assisters are more populous in the thermal soup,  making them  available in large numbers to interact with the DM during the phase of  freeze-out. This results in a boost in the interaction rate, which is evident from  the grey shaded region in figure~\ref{sf:nbolt}. We call this  a \textit{Boltzmann boost}.  In this region or parameter space, the allowed  DM  mass can range from a few to several hundreds of MeV while satisfying the required relic abundance and remaining within the perturbative limit. The region below the $\epsilon=1$ contour in figure \ref{fig:relic}  denotes the relic-density-allowed region for $m_{\phi} > m_{A}~(\epsilon \leq 1)$. For demonstration, we have shown the relic-density-allowed contour for $\epsilon=0.5$ by a green dotted line in figure \ref{fig:relic}.
This  Boltzmann boost is crucial for viable DM masses beyond 20 MeV, while keeping renormalizable perturbativity intact in the theory. As is evident from the black solid lines of figure \ref{fig:bbn-cmb}, as $\epsilon$ decreases, the boost increases, allowing assisted annihilation to   saturate the relic density bound with smaller couplings. This facilitates a  natural $\mathcal{O}(100)$ MeV  thermal DM within the $3\to2$ annihilation   framework without entering the strong coupling regime.

\section{Cosmological Constraints}
\label{sec:cosmo-cons}
The presence of multiple MeV scale  light species can effect the BBN and CMB observations in three ways.
(i) Existence of a particle species with mass less than a few MeV will increase the Hubble expansion rate. This modifies the freeze-out time of neutron-proton interaction, leading to an increased $^4$He abundance. The BBN constraint~\cite{Cyburt:2015mya} from this is shown as the light blue region in figure \ref{fig:relic} and pushes the allowed masses of both DM and the assister to be greater than $\sim 1$ MeV. (ii) Subject to the lifetime,  various decay modes of the assister will also alter BBN and CMB observations. (iii) Additionally a late-time assisted annihilation to photon, electrons and/or neutrinos may alter the neutrino to photon temperature ratio affecting both the BBN and CMB observations \cite{Kolb:1986nf, Serpico:2004nm, Nollett:2013pwa, Nollett:2014lwa}. However, as mentioned earlier, direct annihilation to such light species is precluded in our framework.

\begin{figure*}[t]
\begin{center}
\subfloat[\label{sf:mA10}]{\includegraphics[scale=0.25]{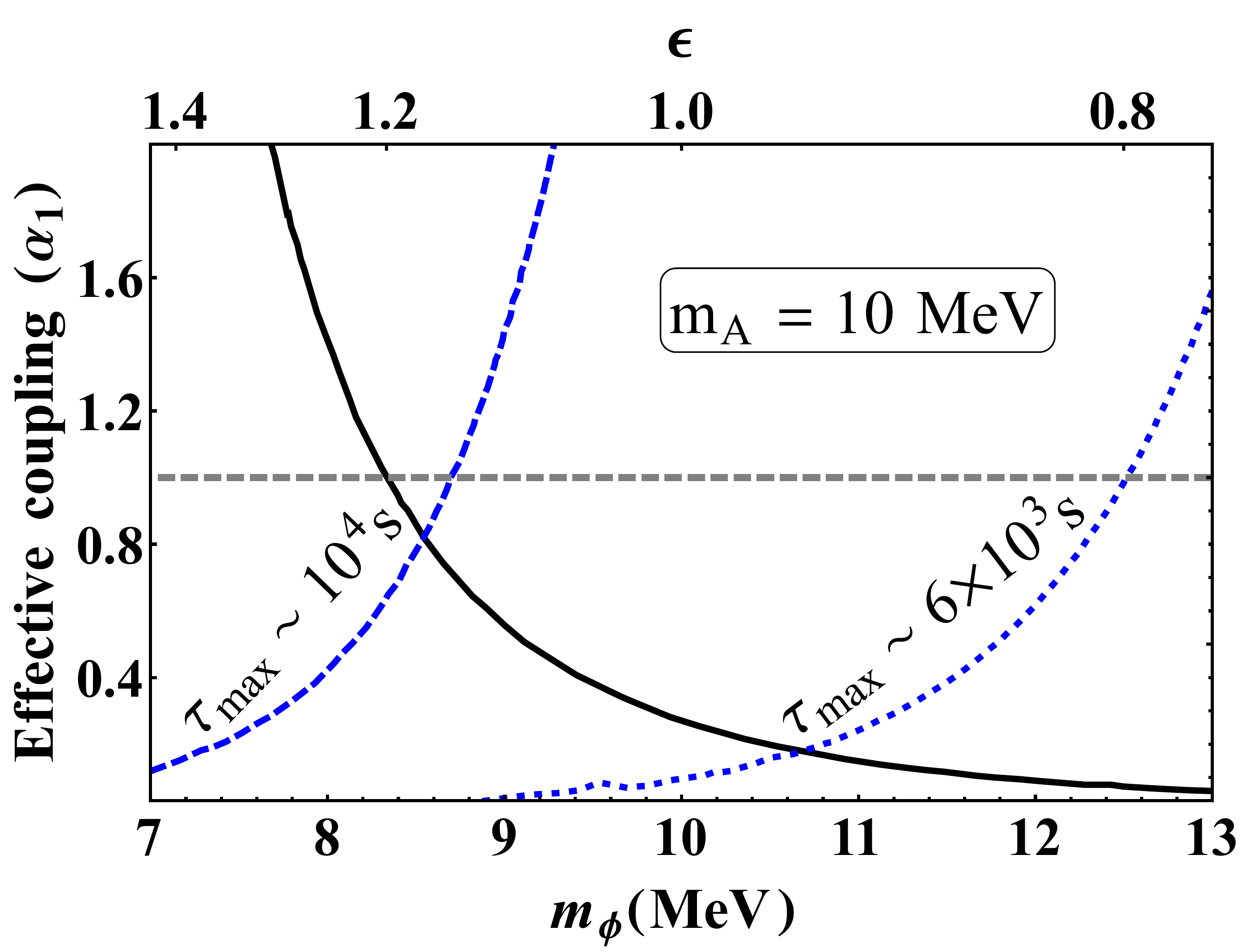}}\quad \quad
\subfloat[\label{sf:mA60}]{\includegraphics[scale=0.25]{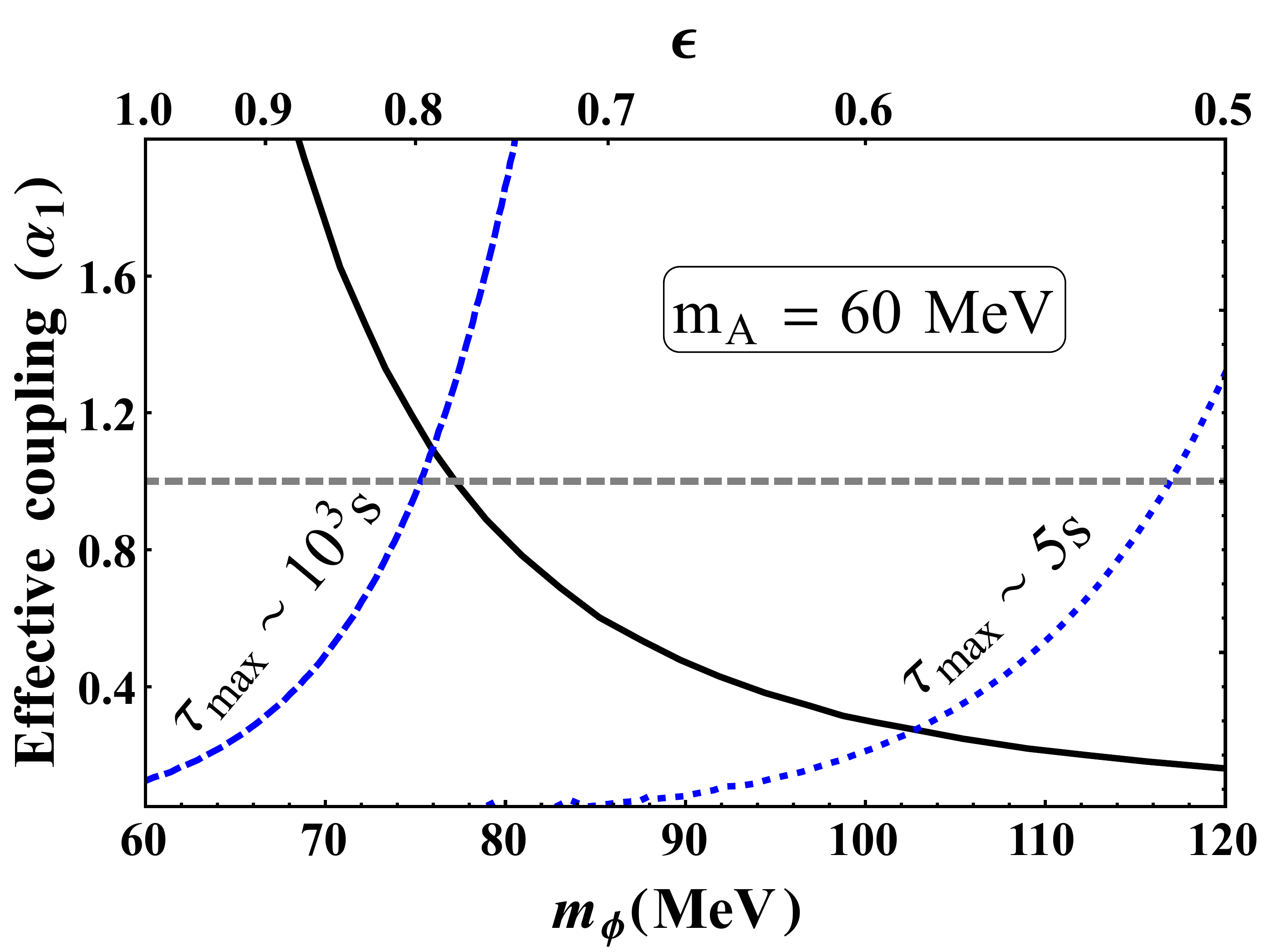}}
\caption{BBN constraint on the lifetime for assister for two benchmark values. (a) For $m_A=10$ MeV the blue dotted and dashed contours show the maximum allowed lifetimes $6 \times 10^3$ and $10^{4}$s with $n_{A}/n_{\gamma} \sim 10^{-5}, 3\times 10^{-7}$ respectively, at the DM freeze-out. (b) For $m_A=60$ MeV, the blue dotted and dashed contours show the maximum allowed lifetime 5 and $10^{3}$s with $n_{A}/n_{\gamma} \sim 2\times 10^{-3}, 10^{-4}$ respectively, at the DM freeze-out. In both of the plots the black solid lines represent the relic-density-allowed values.  The horizontal gray dashed line represents $\alpha_1=1$.}
\label{fig:bbn-cmb}
\end{center}
\end{figure*} 
The BBN or CMB constraints on such decaying species puts an upper bound on the pre-decay yields for a given lifetime.  We will  make the conservative assumption that the DM freeze-out via assisted annihilation is the dominant operative process that keeps the assisters in thermal equilibrium. To constrain such scenario we take the assister yields at the time of DM freeze-out which is calculated at $n_{\phi}^{\rm eq} n_{A}^{\rm eq}\langle \sigma v^{2} \rangle_{\phi \phi A \to \text{SM}~\text{SM}} = H$. Predominant decay of the assister to photons, light leptons or hadrons can perturb the light elements' abundances. Note that coupling to light leptons are inhibited by the corresponding $(g-2)$ measurements \cite{Dey:2016qgf} while for the assister in the mass range of few hundreds of MeV the hadronic decay channels are phase space suppressed. Consequently we  concentrate on photophilic assisters in the rest of the discussion.

Cosmological constraint from BBN on electromagnetically decaying particle has been studied extensively in the literature~\cite{Protheroe:1994dt, Kawasaki:1994sc,Cyburt:2002uv, Jedamzik:2006xz,Poulin:2015opa, Hufnagel:2018bjp, Forestell:2018txr}. We follow the prescription of~\cite{Poulin:2015opa} to numerically solve the integral equation to obtain the non-thermal photon spectra arising from the decay of the assisters. We modify the {\tt AlterBBN v2.0}~\cite{Arbey:2018zfh, Arbey:2011nf} to include photo-dissociation of D, $^3$He, and $^4$He~\cite{Cyburt:2002uv}  along with the  additional contribution to the Hubble parameter to compute the light element abundances in the presence of the light DM-assister states and compare the simulated abundance with the observed value \cite{PhysRevD.98.030001,Cyburt:2015mya}.
In our discussion, we chose two benchmark values for the assister mass, namely, $m_{A} = 10$ and 60 MeV to highlight the BBN constraint on the two regions of Boltzmann boost and suppression as discussed in the previous section. The most stringent constraint for an assister of these masses comes from the D photo-dissociation and not from $^3$He or $^4$He photo-dissociation. Notably, there is no constraint for the assister mass of $10$ MeV from $^3$He or $^4$He photo-dissociation because the photon from the decay of such low mass assister would not be able to overcome the threshold barrier of photo-dissociation of $^3$He and $^4$He \cite{Cyburt:2002uv}. A low mass DM will correspond to higher assisted annihilation cross section (see equation \eqref{se:aaXsection}) and higher DM number density. Thus as the DM mass reduces the assister abundance at DM freeze-out also decreases, leading to easing in constraint from BBN. For an assister mass of $10$ MeV, its abundance $n_{A}/n_{\gamma}$ at the time of DM freeze-out for dotted and dashed blue lines of figure~\ref{sf:mA10} refer to $\sim 10^{-5}, \sim 3\times 10^{-7}$ and the maximum allowed lifetimes from D photo-dissociation are $\tau_{\rm max} \sim 6\times 10^{3}, 10^{4}$s, respectively. Whereas for an assister of mass $60$ MeV its abundance for the dotted and dashed blue lines of figure~\ref{sf:mA60} refer to $\sim 2\times 10^{-3}, \sim 10^{-4}$ and the maximum allowed lifetime from D photo-dissociation are $\tau_{\rm max} \sim 4\times 10^{3}, 5\times 10^{3}$s respectively. 
The decay of the assister after BBN leads to entropy injection which shall change the baryon to photon ratio ($\eta$) \cite{Feng:2003uy}. This effect has been related to the fractional change in the entropy~\cite{Poulin:2015opa}, $\Delta S/S \propto  m_A \sqrt{\tau}\, n_A/n_{\gamma}$, 
where $n_{A}/n_{\gamma}$ is calculated at the DM freeze-out. We set a tolerance of $5 \%$ change in the entropy to be compatible with BBN and CMB. The maximum allowed lifetimes of the assister of mass $10$ MeV  are $10^6\,$ and $10^{9}\,$s for the dotted and dashed contours in figure \ref{sf:mA10}, while for the other one, the maximum allowed lifetime are $5$ and $10^{3}\,$s. 
Finally, the decay of the assister into photon shall reheat the photon bath in comparison to neutrino. This may lead to the reduction of the effective number of relativistic degrees of freedom $N_{\rm eff}$. We have calculated this effect following~\cite{Fradette:2017sdd} and found this constraint to be sub-dominant. The  most stringent limit on the lifetime of an assister of mass $10$  and $60\,$MeV arises from photo disintegration of D and entropy injection, respectively,  which has been depicted in figure \ref{fig:bbn-cmb}.
%

\section{Model Sketch}
\label{sec:toy}
We present a simple scalar completion of the effective framework explored in this article.  Consider a real scalar DM  state $\phi$  stabilized by a discrete $\mathbb{Z}_2$ symmetry. We introduce two additional real  scalar states, the assister $A$ and  a relatively heavier mediator $S,$ both of which can decay to the SM states. The relevant part of the Lagrangian is given by,
\begin{equation}
\mathcal{L} \supset \lambda_{1} \phi^{2}AS 
             + \lambda_{2} S A^{2}
             + \frac{\lambda_{3}}{f}AF^{\mu \nu}F_{\mu \nu},
\label{eq:model-lag}
\end{equation}
\begin{figure}[b]
\begin{center}
\subfloat[\label{sf:toyFeyn1}]{
\includegraphics[scale=0.4]{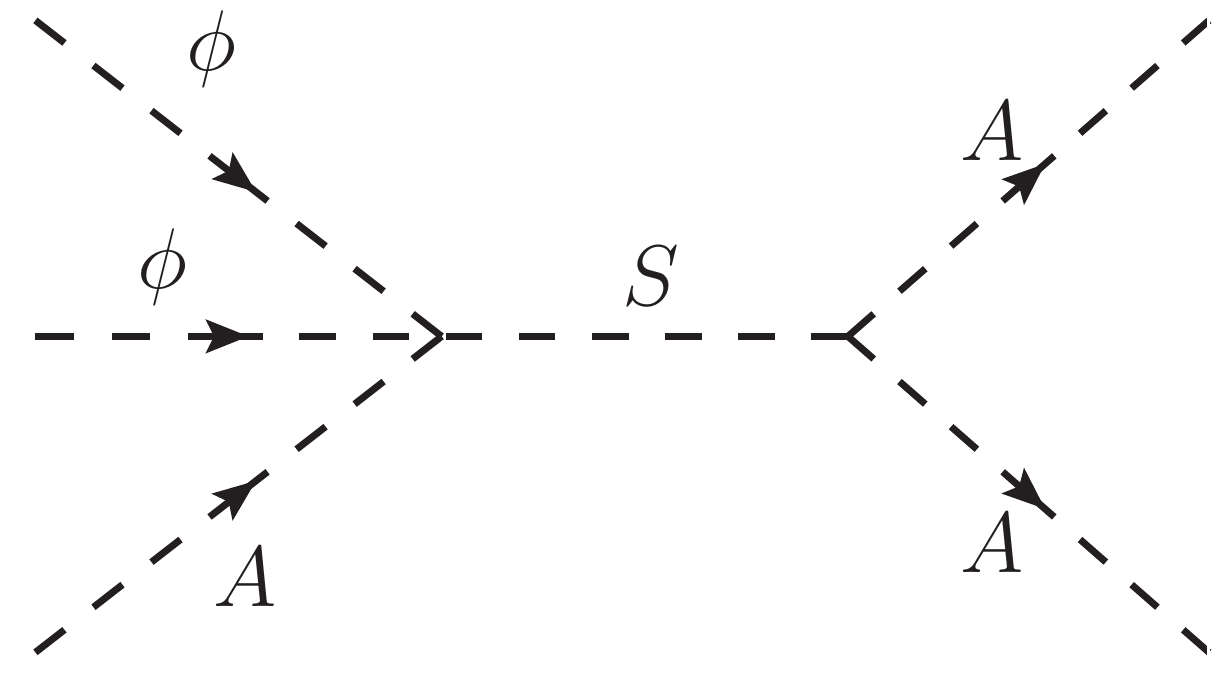}}~~~~
\subfloat[\label{sf:toyFeyn2}]{
\includegraphics[scale=0.4]{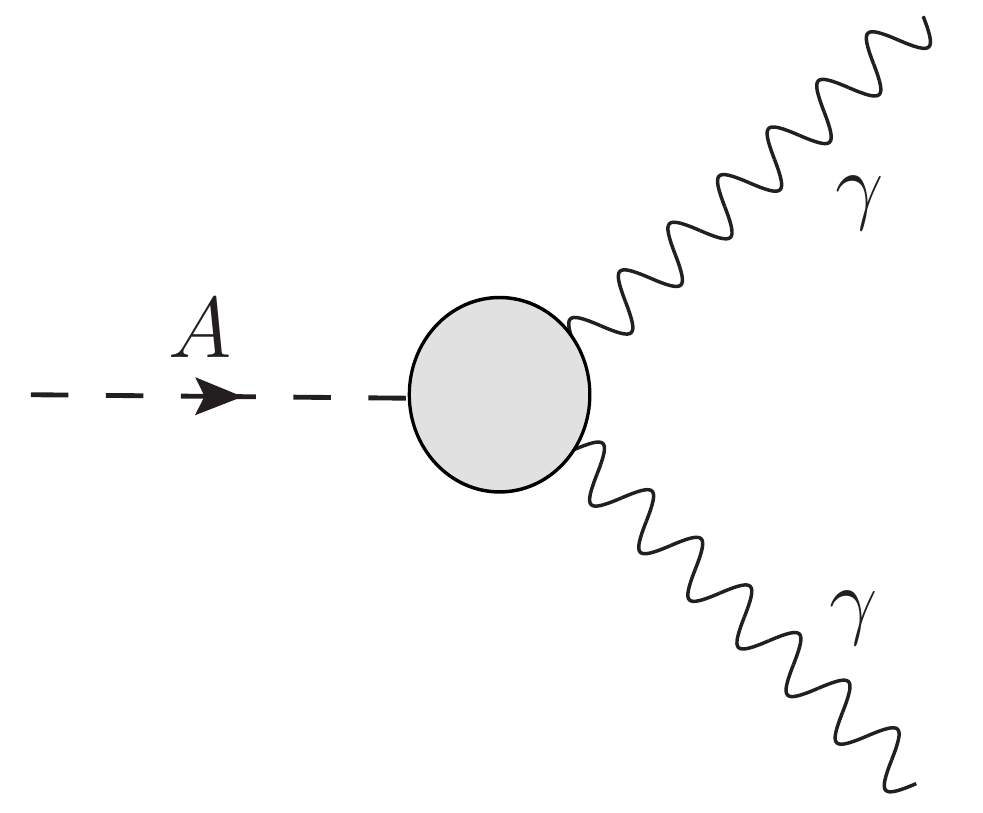}}
\caption{Relevant Feynman diagrams for the toy model. (a) Dominant assisted annihilation channel. (b) The decay of the assister to photons.}
\label{fig:toyFeyn}
\end{center}
\end{figure} 
where $F_{\mu \nu}$ is the usual electromagnetic field strength tensor. The photophilic decay of the assister may  proceed through an SM fermion loop as discussed in \cite{Dey:2016qgf}. A mass hierarchy of  $m_{\phi}\sim m_A \ll m_S$ ensures that the $3\to 2$  assisted annihilation process ($ \phi \phi A \rightarrow A\,A$)  through the  $S$ mediation as shown in figure~\ref{fig:toyFeyn} will control DM freeze-out. For the   associated $2\to 2$ process $\phi \phi \leftrightarrow A S,$  the forward rate  has a large phase space suppression, while the backward process is quenched by a large Boltzmann factor  arising due to the  mass difference between the mediator  and the assister. Note that, for $\epsilon < 2/3$ the annihilation channel $\phi \phi \to AAA$ would open up and start dominating over the assisted annihilation mechanism to drive freeze-out. Whereas for $\epsilon >2$ assisted annihilation process $\phi \phi A \rightarrow A\,A$ becomes ineffective due to phase space suppression. Since the $n$-body annihilation rate of a process is  proportional to exp$(-\sum_{i=1}^{n} m_i/T)$, for $\epsilon <1$  contribution of $A A A \to \phi \phi $ may become significant. In the window of $1 < \epsilon < 2$ assisted annihilation remains the dominant processes to control the freeze-out. As an explicit benchmark point, we find that for the DM and the assister around $10$ MeV, a mediator mass $m_S \gtrsim 40$ MeV renders contribution of the $\phi \phi A \rightarrow A\,A$ process  numerically significant. For these masses, the required relic density can be easily obtained with perturbative  couplings at $\lambda_{1} =\lambda_{2}/m_{\phi} \sim 2.$ The coupling $\lambda_3/f$ can be independently set to align the  decay width of the assister with the cosmological constraints discussed in the previous section. 
A complimentary  search strategy  for such photophilic assister is the fixed target experiments.  The assisters may be produced (i) by the Primakoff-like process in E137 \cite{Bjorken:1988as}, FASER2 \cite{Feng:2018noy}, NA62 \cite{Dobrich:2015jyk}, SeaQuest \cite{Berlin:2018pwi}, P\textsc{rim}E\textsc{x} \cite{Aloni:2019ruo},  and SHiP \cite{Alekhin:2015byh}, (ii) from the $e^{+}e^{-}$ annihilation via the axion-like-particle-strahlung process and photon fusion in Belle-II \cite{Dolan:2017osp},  (iii) from the exotic decays of light mesons in FASER2 \cite{Feng:2018noy}. We have adopted the results of \cite{ Alekhin:2015byh, Dobrich:2015jyk, Dolan:2017osp, Feng:2018noy, Berlin:2018pwi, Aloni:2019ruo} to obtain constraints on the parameter $\lambda_3/f$ for the two benchmark masses of the assister depicted in figure \ref{fig:bbn-cmb}, and the corresponding limit has been given in  table \ref{table}. For comparison we list the  corresponding  BBN exclusion limit. The quoted numbers represent the most stringent BBN exclusion limit on $\lambda_3/f$ in the parameter space $\epsilon>2/3$. Interestingly, the limits from fixed target experiments are complimentary to the ones from BBN.  However, as demonstrated in table~\ref{table}, the resolution of these present and proposed experiments would leave a window of several orders of magnitude of allowed parameter space.  
\begin{table}[t]
\centering  
\begin{tabular}{|c|c|c|c|}
\hline  
 \multirow{3}{3em}{ ~$m_A$ (MeV)}       &  \multicolumn{2}{c|}{Fixed target experiments } &\multirow{3}{7.5em}{$\frac{\lambda_3 }{f} $  ($10^{-5}$GeV$^{-1}$)  BBN exclusion} \\ \cline{2 -3}
        &  Experiment          &$\frac{\lambda_3}{f} $  ($10^{-5}$GeV$^{-1}$)    & \\
 & name        & Exclusion limit      &  \\ 
\hline \hline
\multirow{8}*{60}   

& P\textsc{rim}E\textsc{x} \cite{Aloni:2019ruo}   & $ \gtrsim 20$   &  \\ \cline{2 -3}

& G\textsc{lue}X \cite{Aloni:2019ruo}   & $ \gtrsim 7.3$   &  \\ \cline{2 -3}

& E137 \cite{Bjorken:1988as}   & $0.033 - 7.3  $   &  \\ \cline{2 -3}

& Belle-II \cite{Dolan:2017osp}     & $0.9 - 20$  &  \\ \cline{2 -3}

& SHiP \cite{Alekhin:2015byh}       & $0.067 - 30$      &$\lesssim 8.8 \times 10^{-6}$  \\ \cline{2 -3}

& FASER2 \cite{Feng:2018noy}        & $0.074 - 60$    & \\ \cline{2 -3}

& SeaQuest \cite{Berlin:2018pwi}    & $0.04 - 20$       & \\ \cline{2 -3}

& NA62  \cite{Dobrich:2015jyk}     & $0.41 - 20$      & \\ \hline

\multirow{8}*{10}  
& P\textsc{rim}E\textsc{x} \cite{Aloni:2019ruo}   & -  &  \\ \cline{2 -3}

& G\textsc{lue}X \cite{Aloni:2019ruo}   & -   &  \\ \cline{2 -3}
   
& E137 \cite{Bjorken:1988as}  & $0.14 - 400$  &  \\ \cline{2 -3}

& Belle-II \cite{Dolan:2017osp}    & $0.9 - 1000$    &  \\ \cline{2 -3}

& SHiP \cite{Alekhin:2015byh}      & $0.44 - 1000$       &$\lesssim 5.0 \times 10^{-6}$  \\ \cline{2 -3}

& FASER2 \cite{Feng:2018noy}       & $4.4 - $  & \\ \cline{2 -3}

& SeaQuest \cite{Berlin:2018pwi}   & $0.2 - 800$       & \\ \cline{2 -3}

& NA62  \cite{Dobrich:2015jyk}    & $2- 1000$      & \\ \hline
\end{tabular}
\renewcommand{\arraystretch}{0.5}
\caption{Exclusion region of the parameter $\lambda_3/f$ for different fixed target experiments for assister mass of $60$ and $10$ MeV. For comparison the lower bound from cosmological observation has also been shown in the fourth column.}
\label{table}
\end{table}
This can be viewed as a subset of a theory with a $\mathbb{Z}_2 \times \mathbb{Z'}_2$ symmetry under which the $\phi ~(A,S)$ have charges $[-,+] ~([+,-])$ while the SM states are all $[+,+].$ In a different sector, the $\mathbb{Z'}_2$ symmetry is explicitly broken  leading to  the decay of the assister and the mediator to SM states. Admittedly in most regions of parameter space of such a theory the $2 \to 2$ annihilation dominates, yielding the usual GeV scale DM. Interestingly, in a corner of parameter space of the theory, where the operators in the Lagrangian given in equation~\eqref{eq:model-lag} dominate over every other term and the mass hierarchy is given by  $m_{\phi}\sim m_A \ll m_S,$ the assisted annihilation  mechanism  is operative in setting  the DM relic density. This is a complementary region of allowed parameter space  of the full theory with an  assisted annihilation driven scalar LDM in the MeV scale.  
%

\section{Conclusion}
\label{sec:concl}
The mechanism of assisted annihilation accommodates appropriate freeze-out for thermal light dark matter with the help of the assisters in the initial state. In this article, we show that when assister(s) are lighter than the DM a Boltzmann boost can significantly enhance assisted annihilation. This augmentation is an atypical feature of this framework that considerably relaxes the upper limit on DM masses for  the $3\to2$ annihilation dominated paradigm to $\mathcal{O}(100)$ MeV without any strong couplings. We demonstrate that such a scenario is cosmologically viable from BBN and CMB observations while saturating the relic density bound. This natural LDM framework can be embedded in simple particle physics models and may provide a handle for addressing small-scale structure formation issues through appropriate self-interactions.

\acknowledgments 
The research of UKD is supported by the Ministry of Science, ICT \& Future Planning of Korea, the Pohang City Government, and the Gyeongsangbuk-do Provincial Government. TNM would like to thank MHRD, Government of India for research fellowship. TSR is partially supported by the Department of Science and Technology,  Government of India, under the Grant Agreement No. IFA13-PH-74 (INSPIRE Faculty Award)

\bibliographystyle{JHEP}
\bibliography{ref.bib}

\end{document}